\documentclass[journal]{IEEEtran}
\usepackage[T1]{fontenc}
\usepackage{amsmath,amssymb,amsfonts}
\usepackage{graphicx}
\usepackage{cite}
\usepackage{booktabs}
\usepackage{algorithm}
\usepackage{algpseudocode}
\usepackage{xcolor}
\usepackage[hidelinks]{hyperref}
\usepackage{subfig}
\newcommand{\tX}{\tilde{X}}

\begin{document}

\title{Interference-Free In-Symbol Pilots for High-Mobility DFT-s-OFDM Uplink}

\author{Ayoub~Ammar~Boudjelal
and~H\"{u}seyin~Arslan,~\IEEEmembership{Fellow,~IEEE}%
\thanks{The authors are with the Department of Electrical and Electronics
Engineering, Istanbul Medipol University, Istanbul 34810, T\"{u}rkiye
(e-mail: ayoub.ammar@std.medipol.edu.tr;
huseyinarslan@medipol.edu.tr).}}

\AddToHook{shipout/foreground}{\put(0.13\paperwidth,-0.035\paperheight){\parbox[t]{0.74\paperwidth}{\footnotesize\centering This work has been submitted to the IEEE Transactions on Communications for possible publication. Copyright may be transferred without notice, after which this version may no longer be accessible.}}}
\maketitle

\begin{abstract}
The 5G New Radio (NR) uplink uses discrete
Fourier transform spread orthogonal frequency-division multiplexing (DFT-s-OFDM) for its low peak-to-average power ratio
(PAPR). Its demodulation reference signal (DMRS) occupies
whole symbols, although channel estimation only needs channel
samples. Since existing designs cannot place pilots inside a data
symbol without interference on the data, or on both data and
pilots, NR keeps whole DMRS symbols, which wastes spectral
efficiency (SE), and adds more of them as the channel varies
faster, lowering the SE further. We develop a technique that
places pilots free of data interference inside the data symbols,
for frequency-domain channel estimation in every symbol. Using
the Zak transform, we show that leaving a set of equally spaced
pre-DFT samples empty frees one column of the Zak grid of
the DFT output. The data values of the pilot subcarriers are
relocated into this column by adding a constant over a few
subcarrier combs, which preserves the energy and keeps the
NR DFT sizes. The receiver undoes it linearly with comb sums,
without decisions or side information, and a symbol loses as few
as one data sample beyond its pilots. Since the pilots raise the
PAPR, we keep them few per symbol and hop their positions
across symbols, so that a two-dimensional estimator over the slot
resolves long delay spreads at high Doppler. On 3rd Generation Partnership Project (3GPP) channels,
pilots in every symbol achieve a higher effective SE than all
NR DMRS configurations from pedestrian to high-speed-train
velocities, at a PAPR close to plain DFT-s-OFDM.
\end{abstract}


\section{Introduction}
\label{sec:intro}

\IEEEPARstart{L}{ong} term evolution (LTE) and 5G new radio (NR) rely on discrete Fourier transform spread orthogonal frequency-division multiplexing (DFT-s-OFDM), also known as single-carrier frequency-division multiple access (SC-FDMA), for coverage-limited uplink transmission~\cite{myung2006,dahlman2020,3gpp38211}. By spreading the data symbols with a DFT, DFT-s-OFDM produces a signal whose envelope follows the constellation, so its peak-to-average power ratio (PAPR) is well below that of orthogonal frequency-division multiplexing (OFDM)~\cite{myung2006,wunder2013}. This property is critical at the cell edge, where a user equipment (UE) transmits at its maximum power and every decibel of PAPR translates into a decibel of power-amplifier back-off, and hence of lost link budget. At the same time, these UEs are increasingly mobile: high-speed trains, vehicle-to-everything links, aerial UEs, and non-terrestrial access all combine a power-limited uplink with a channel that varies within a slot~\cite{wu2016}. Indeed, the waveforms proposed for 6G are designed precisely for such doubly dispersive channels~\cite{zhang2026,rou2024}.

In NR, channel estimation for the transform-precoded physical uplink shared channel (PUSCH) is performed with the demodulation reference signal (DMRS), which occupies whole OFDM symbols with a low-PAPR sequence and carries no data~\cite{3gpp38211,3gpp38214}. As the Doppler spread grows, additional DMRS symbols are configured~\cite{3gpp38211}. This approach faces two limitations. First, every additional DMRS symbol removes a symbol of data. Second, the channel is still sampled in only a few symbols and must be interpolated in between, which becomes inaccurate as mobility increases. Both limitations stem from a single constraint: in DFT-s-OFDM, a pilot cannot share a symbol with data without raising the PAPR, losing data, or suffering data interference. This paper relaxes that constraint. Specifically, we seek pilots inside the data symbols that (i) are free of data interference, so that only noise and inter-carrier interference (ICI) remain on them, (ii) cost about as many data samples as there are pilots, (iii) do not significantly increase the PAPR, (iv) are compatible with the NR DFT sizes and transmitter chain, (v) have negligible complexity, and (vi) need no side information. To the best of our knowledge, no existing design satisfies all of these requirements simultaneously, as discussed next.

\subsection{Related Literature}
Apart from the DMRS, the other NR uplink reference signals do not fill this gap~\cite{3gpp38211,dahlman2020}. The sounding reference signal (SRS) occupies a frequency comb in its own symbols; the phase-tracking reference signal (PT-RS) is a set of pre-DFT chunks spread over all subcarriers and tracks only the common phase; and DMRS bundling~\cite{3gpp38830,3gpp38214} estimates the channel jointly across slots, which benefits slowly varying channels only. For OFDM, pilot placement in time and frequency has been studied extensively~\cite{coleri2002,choi2005}, with the two-dimensional Wiener filter~\cite{hoeher1997,li1998} serving as the reference estimator. Scattered pilots whose frequency positions shift from symbol to symbol, as in digital video broadcasting~\cite{dvbt}, sample the time-frequency plane with fewer pilots than pilots repeated at the same subcarriers, and the pilot pattern can be optimized for the channel statistics~\cite{choi2005}. When the channel varies within a symbol, the ICI~\cite{wang2006} must additionally be modeled or cancelled. These techniques, however, do not carry over to DFT-s-OFDM: since every subcarrier carries a combination of all data symbols, no subcarrier is free to host a scattered pilot, and the estimator remains confined to the DMRS symbols. Data-aided receivers can refine the estimate between DMRS symbols, but at the cost of iterations and error propagation from wrong decisions; they are complementary to any pilot design and are not considered here.

Several approaches can place pilots inside a data symbol, but each violates at least one of the above requirements. The most direct one is frequency-division multiplexing (FDM), where the data are spread by a smaller DFT and the pilots occupy the remaining subcarriers. The data and pilots then no longer form a single-carrier signal, which raises the PAPR, and the smaller DFT size is in general not an NR size~\cite{3gpp38211}. Puncturing, which overwrites subcarriers of the DFT output with pilots, instead erases part of every data symbol. Superimposed training~\cite{tugnait2003} adds pilots to the data at no resource cost, but the pilots and data interfere, so the estimate saturates at high signal-to-noise ratio (SNR) unless the data are cancelled iteratively. Data-dependent superimposed training (DDST)~\cite{ghogho2005} removes this interference by subtracting the data component on the pilot frequencies; in DFT-s-OFDM, however, it discards part of every data symbol in a way that a linear receiver cannot recover. Similarly, a DMRS placed before the DFT inside the data symbols, guarded by a cyclic prefix and suffix, removes the need for time interpolation for high-mobility users~\cite{gudimitla2024}. Such training, however, is spread over all subcarriers and overlaps with the data in frequency, so the channel cannot be read on any subcarrier without first removing the data. Our construction also leaves pre-DFT samples empty, but uses them for a different purpose: to free pilot subcarriers. Orthogonal time frequency space (OTFS) modulation~\cite{hadani2017,wei2021} places data and pilots on a delay-Doppler grid, where an embedded pilot surrounded by guard cells provides an interference-free estimate of a doubly selective channel~\cite{raviteja2019,raviteja2018}. Its Zak formulation~\cite{mohammed2022,mohammed2023,jayachandran2025} relies on the discrete Zak transform~\cite{bolcskei1997}, which maps a signal onto a two-dimensional grid. Affine frequency-division multiplexing~\cite{bemani2023} achieves similar properties with chirp subcarriers and is positioned as an extension of the OFDM transceiver~\cite{yin2025}. Nevertheless, these waveforms replace DFT-s-OFDM, require a new receiver, and in general exhibit a higher PAPR~\cite{surabhi2019,rou2024,mehrotra2025}. Moreover, their gain over OFDM with equally tailored pilots and estimators depends on the channel and receiver assumptions~\cite{gaudio2022}, and the guard region of the embedded pilot grows with the delay and Doppler spreads at the expense of data cells. In this work, we borrow the Zak grid, but only to locate free space inside an unchanged DFT-s-OFDM symbol. The discrete Zak transform is the polyphase decomposition of a sequence followed by a DFT across the polyphase components~\cite{bolcskei1997}. A special case of this structure is well known in DFT-s-OFDM: with interleaved subcarrier mapping, a single comb carries a repeated, phase-rotated copy of the data block in time~\cite{myung2006}. The identity underlying our construction, namely that the pre-DFT samples at the multiples of a stride are all zero if and only if every subcarrier comb of the complementary spacing sums to zero, follows from this polyphase structure. To our knowledge, however, it has not been exploited to free pilot subcarriers inside a DFT-s-OFDM data symbol without discarding data.

Among classical PAPR reduction methods~\cite{wunder2013}, tone reservation~\cite{krongold2004} cancels peaks on reserved subcarriers without side information, but in DFT-s-OFDM it takes subcarriers away from the data~\cite{gokceli2021}. NR lowers the PAPR of DFT-s-OFDM with $\pi/2$ binary phase-shift keying (BPSK)~\cite{3gpp38211} and with frequency-domain spectrum shaping (FDSS), which the UE may apply within the flatness requirements of~\cite{3gpp381011} and which spectral extension can strengthen~\cite{peruga2021}; such enhancements are being unified into DFT-s-OFDM-based 6G waveforms~\cite{liu2023}. Since these techniques act on the data, they are compatible with the proposed scheme. Furthermore, in the proposed scheme, the empty-column cells that hold no relocated data can carry a peak-cancelling signal, turning PAPR reduction into an option traded for data samples.

\begin{figure*}[!t]
\centering
\includegraphics[width=0.92\textwidth]{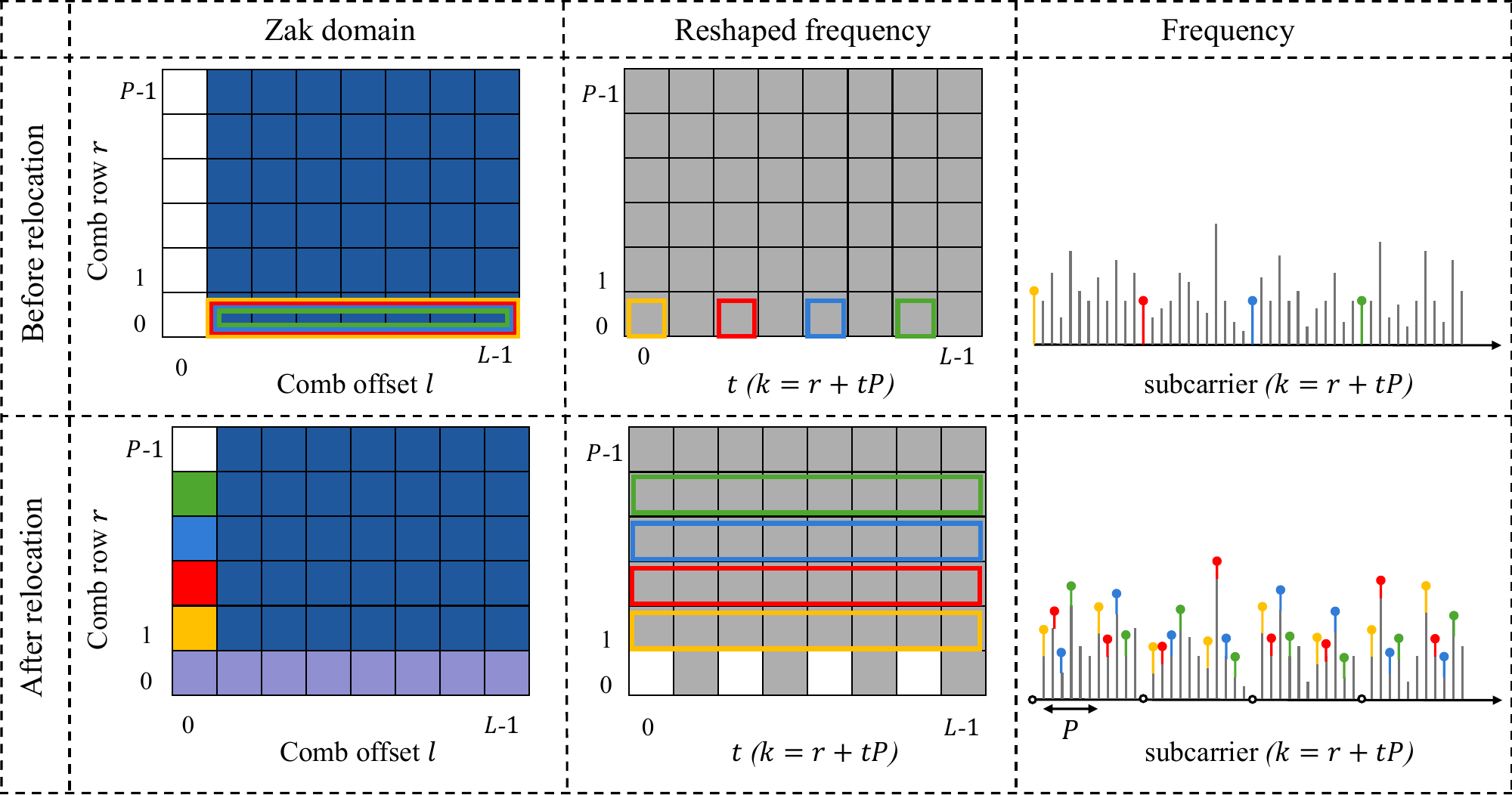}
\caption{Zak grid, subcarriers arranged by comb, and subcarriers before (top) and after (bottom) the relocation ($N_\mathrm{p}=4$).}
\label{fig:zak}
\end{figure*}

\subsection{Motivation and Contributions}

In summary, existing designs either keep the pilots out of the data symbols (NR DMRS, SRS, DMRS bundling), spread them over all subcarriers (PT-RS, ZT/UW, pre-DFT DMRS), sacrifice PAPR or NR compatibility, or lose data or suffer data interference (puncturing, superimposed training, DDST). None of them places pilots free of data interference inside a DFT-s-OFDM data symbol at a data cost comparable to the number of pilots, while keeping the NR DFT size and only a small PAPR increase. This motivates the scheme proposed in this paper, which exploits the Zak structure of the DFT-s-OFDM symbol to create such pilots within the existing transmitter chain. To assess it fairly, we evaluate the proposed scheme and the NR DMRS configurations with the same two-dimensional estimator on 3rd Generation Partnership Project (3GPP) tapped-delay-line (TDL) channels, including the ICI and inter-symbol interference that these channels produce. The main contributions of this paper are as follows:
\begin{itemize}
\item We develop a technique that places pilots free of data interference inside a DFT-s-OFDM data symbol, enabling frequency-domain channel estimation within the data symbols. Using the Zak transform, we show that leaving one stride class of $P$ pre-DFT samples empty frees one column of the $P \times L$ Zak grid of the DFT output, whose rows are subcarrier combs. We then relocate the data values of the $N_\mathrm{p}$ pilot subcarriers into this column by adding a constant over $N_\mathrm{p}$ combs. The relocation preserves the energy, keeps the NR DFT size and transmitter chain, costs $P \ge N_\mathrm{p}+1$ data samples and fewer additions than the DFT size, and is undone linearly with comb sums, without decisions or side information.

\item We propose to hop the pilot positions from symbol to symbol, so that the channel can be estimated with few pilots per symbol: the pilots of consecutive symbols interleave in frequency and are combined by a two-dimensional (2-D) linear minimum mean square error (LMMSE) filter over the slot. Keeping the number of pilots per symbol small is essential, since the pilots and the relocated values raise the PAPR; accordingly, we derive the delay and Doppler conditions that the hopping pattern must satisfy.

\item We employ Zadoff-Chu (ZC) pilots, whose flat DFT spreads the pilot energy evenly over the grid, and we use the free cells of the empty column for optional peak cancellation without side information, which trades data samples for PAPR.

\item We evaluate the PAPR, channel estimation accuracy, bit error rate (BER), and effective spectral efficiency against the NR DMRS configurations of PUSCH mapping type A on 3GPP channels, ranging from pedestrian to train speeds.
\end{itemize}

\subsection{Organization}
The rest of the paper is organized as follows. Section~\ref{sec:tx} follows a symbol through the transmitter, from the data mapping and the DFT to the Zak grid, the relocation, the pilots and the optional peak cancellation. Section~\ref{sec:ch} gives the channel model, Section~\ref{sec:rx} derives the receiver, and Section~\ref{sec:nr} discusses the overhead, the complexity and the integration into NR. Section~\ref{sec:sim} presents the simulation results, and Section~\ref{sec:Conc} concludes. Throughout the paper, $j=\sqrt{-1}$, square brackets denote discrete indices, $(\cdot)^*$ denotes complex conjugation, $\bmod$ denotes the remainder of integer division, $\delta[\cdot]$ is the Kronecker delta, $\mathbb{E}[\cdot]$ is the expectation, $|\mathcal{S}|$ is the cardinality of $\mathcal{S}$, $\leftarrow$ denotes assignment, $\mathcal{CN}(0,\sigma^2)$ is the circularly symmetric complex Gaussian distribution, and a tilde marks quantities after the relocation.

\begin{figure*}[!t]
\centering
\includegraphics[width=\textwidth]{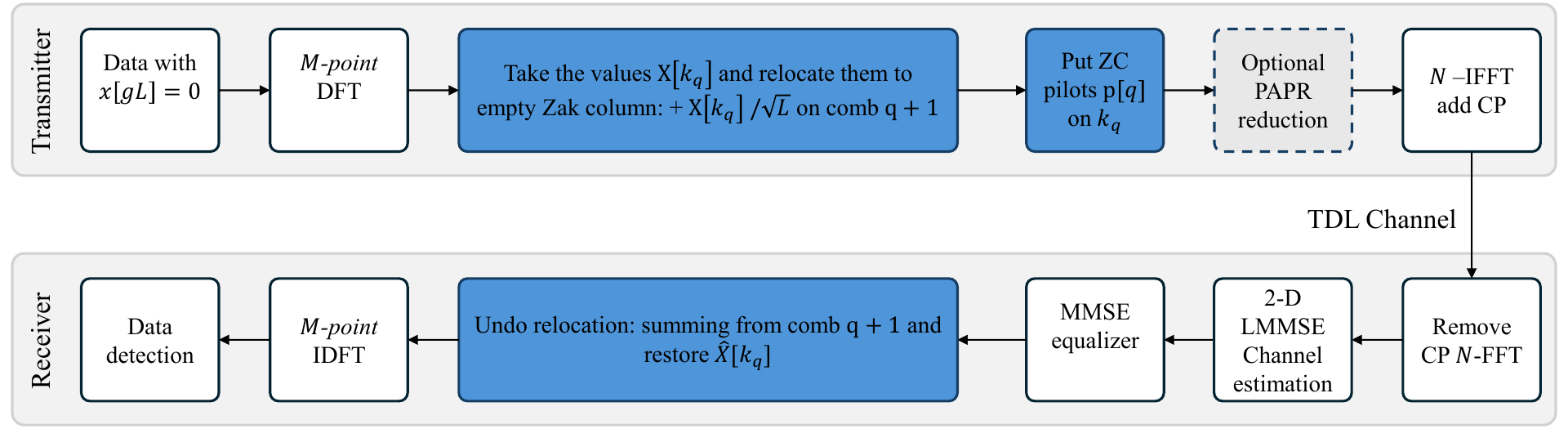}
\caption{Signal chain of the proposed scheme.}
\label{fig:block}
\end{figure*}

\section{Transmitter}
\label{sec:tx}

This section follows symbol $m=0,\dots,N_\mathrm{s}-1$ of a slot of $N_\mathrm{s}$ DFT-s-OFDM symbols. The symbol index $m$ appears only in the pilot positions, the only quantity that changes from symbol to symbol.

\subsection{Data Mapping}
The UE transmits a block of $M$ pre-DFT samples, where $M$ is an NR DFT size, i.e., $M=12N_\mathrm{RB}$, where the number of resource blocks $N_\mathrm{RB}$ has only the prime factors 2, 3 and 5~\cite{3gpp38211}. We factor $M=PL$ and index the samples as $x[l+gL]$, with the time offset $l=0,\dots,L-1$ and the group $g=0,\dots,P-1$; the $P$ samples with the same $l$, spaced $L$ apart, form a stride class. The samples with $l=0$ are empty:
\begin{equation}
x[l+gL]=\begin{cases}0, & l=0,\\ \text{data}, & l\neq0,\end{cases}
\label{eq:map}
\end{equation}
with the unit-power data symbols $d[\cdot]$ ordered by $l+gL$.

\subsection{DFT Spreading}
The $M$-point DFT gives the subcarrier values
\begin{equation}
X[k]=\frac{1}{\sqrt{M}}\sum_{g=0}^{P-1}\sum_{l=0}^{L-1}x[l+gL]\,e^{-j2\pi k(l+gL)/M},
\label{eq:dft}
\end{equation}
for $k=0,\dots,M-1$, and its inverse is
\begin{equation}
x[l+gL]=\frac{1}{\sqrt{M}}\sum_{k=0}^{M-1}X[k]\,e^{j2\pi k(l+gL)/M}.
\label{eq:idft}
\end{equation}
We write $k=r+tP$, with the comb index $r=0,\dots,P-1$ and the comb position $t=0,\dots,L-1$; the $L$ subcarriers with the same $r$, spaced $P$ apart, form comb $r$.

\subsection{Zak Transform}
The Zak transform arranges the subcarriers on a $P\times L$ grid whose row $r$ is the $L$-point inverse DFT (IDFT) of comb $r$:
\begin{equation}
Z[r,l]=\frac{1}{\sqrt{L}}\sum_{t=0}^{L-1}X[r+tP]\,e^{j2\pi lt/L}.
\label{eq:zak}
\end{equation}
In~\eqref{eq:idft} with $k=r+tP$, the exponent is $2\pi(rl/M+rg/P+tl/L+tg)$, and the sum over $t$ is $\sqrt{L}\,Z[r,l]$ by~\eqref{eq:zak}. Hence
\begin{equation}
x[l+gL]=\frac{1}{\sqrt{P}}\sum_{r=0}^{P-1}Z[r,l]\,e^{j2\pi r(l+gL)/M},
\label{eq:col}
\end{equation}
i.e.\ the samples with time offset $l$ depend only on column $l$ of the grid and, up to the phases $e^{j2\pi rl/M}$, are its $P$-point IDFT. As this IDFT is invertible, the samples with time offset 0 are all zero, as in~\eqref{eq:map}, if and only if column 0 is empty, and then, by~\eqref{eq:zak} with $l=0$, every comb sums to zero:
\begin{equation}
\sum_{t=0}^{L-1}X[r+tP]=\sqrt{L}\,Z[r,0]=0\quad\text{for every } r.
\label{eq:empty}
\end{equation}
The inverse of~\eqref{eq:zak}, which recovers comb $r$ from row $r$,
\begin{equation}
X[r+tP]=\frac{1}{\sqrt{L}}\sum_{l=0}^{L-1}Z[r,l]\,e^{-j2\pi lt/L},
\label{eq:izak}
\end{equation}
shows that a value $a$ placed in the empty cell $(r,0)$ adds the constant $a/\sqrt{L}$ to every subcarrier of comb $r$, and by~\eqref{eq:col} it changes only the samples with time offset 0. The $P$ cells of column 0 are therefore free resources that do not interfere with the data. Fig.~\ref{fig:zak} shows the grid.

\subsection{Pilots}
In symbol $m$, the $N_\mathrm{p}$ pilots occupy comb 0 at the subcarriers
\begin{equation}
k_q=\Big(\frac{qL}{N_\mathrm{p}}+(m\bmod N_\mathrm{h})\frac{L}{N_\mathrm{p}N_\mathrm{h}}\Big)P,\quad q=0,\dots,N_\mathrm{p}-1,
\label{eq:bins}
\end{equation}
spaced $M/N_\mathrm{p}$ apart. Over $N_\mathrm{h}$ consecutive symbols the positions hop to $N_\mathrm{p}N_\mathrm{h}$ different subcarriers spaced $M/(N_\mathrm{p}N_\mathrm{h})$ apart, at no cost, since all positions stay on comb 0; $N_\mathrm{h}=1$ keeps the pilots on the same subcarriers. This requires $N_\mathrm{p}N_\mathrm{h}$ to divide $L$. The data values $X[k_q]$ that occupy these subcarriers before the pilots are inserted are relocated into the cells $(q+1,0)$, on combs $1,\dots,N_\mathrm{p}$, which requires $P\ge N_\mathrm{p}+1$. For a given $M$, $N_\mathrm{p}$ and $N_\mathrm{h}$, the smallest $P$ that divides $M$, with $N_\mathrm{p}N_\mathrm{h}$ dividing $M/P$ and $P\ge N_\mathrm{p}+1$, minimizes the number of empty samples. In the best case $P=N_\mathrm{p}+1$ and the pilots cost one data sample more than their number; otherwise the factorization of $M$ forces a larger $P$, and the further empty samples can carry peak cancellation.

The pilots are a ZC sequence~\cite{chu1972},
\begin{equation}
p[q]=e^{-j\pi u\,q\,(q+b)/N_\mathrm{p}},\quad b=N_\mathrm{p}\bmod 2,
\label{eq:zc}
\end{equation}
with root $u$ coprime to $N_\mathrm{p}$ and unit magnitude, i.e.\ the power of a data subcarrier. By~\eqref{eq:zak} and~\eqref{eq:bins}, they contribute to row 0
\begin{equation}
\frac{1}{\sqrt{L}}\,e^{j2\pi l(m\bmod N_\mathrm{h})/(N_\mathrm{p}N_\mathrm{h})}\sum_{q=0}^{N_\mathrm{p}-1}p[q]\,e^{j2\pi lq/N_\mathrm{p}},
\label{eq:pilotrow}
\end{equation}
i.e.\ $\sqrt{N_\mathrm{p}/L}$ times the unitary $N_\mathrm{p}$-point IDFT of the pilots, periodic in $l$ with period $N_\mathrm{p}$ and rotated by the hopping phase. A ZC sequence keeps a constant magnitude under the DFT, so~\eqref{eq:pilotrow} has the magnitude $\sqrt{N_\mathrm{p}/L}$ at every time offset: the pilot energy is spread evenly over row 0 and does not add coherently to a few data samples. 

\subsection{Relocation}
The relocation adds each data value of a pilot subcarrier, divided by $\sqrt{L}$, to every subcarrier of one comb, and writes the pilot on the pilot subcarrier:
\begin{equation}
\tX[k]=\begin{cases}p[q], & k=k_q,\\[2pt] X[k]+\dfrac{X[k_q]}{\sqrt{L}}, & k\bmod P=q+1,\\[4pt] X[k], & \text{otherwise}.\end{cases}
\label{eq:reloc}
\end{equation}
No Zak transform is computed. To see what~\eqref{eq:reloc} does on the grid, insert it into~\eqref{eq:zak}: on comb $q+1$, the added constant contributes $X[k_q]\,\frac{1}{L}\sum_t e^{j2\pi lt/L}=X[k_q]\,\delta[l]$, so with~\eqref{eq:empty}
\begin{equation}
\tilde Z[r,l]=\begin{cases}X[k_{r-1}], & l=0,\ r=1,\dots,N_\mathrm{p},\\ Z[r,l], & \text{otherwise, } r\ge1,\end{cases}
\label{eq:zakreloc}
\end{equation}
i.e.\ each relocated value fills one empty cell and the data cells of combs $1,\dots,P-1$ are unchanged. By~\eqref{eq:empty}, the cross terms between comb $q+1$ and the added constant vanish, so the energy added to comb $q+1$ is $|X[k_q]|^2$, exactly the energy removed from subcarrier $k_q$:
\begin{equation}
\sum_{k=0}^{M-1}|\tX[k]|^2=\sum_{k=0}^{M-1}|X[k]|^2+\sum_{q=0}^{N_\mathrm{p}-1}|p[q]|^2.
\label{eq:energy}
\end{equation}
In time, by~\eqref{eq:col} with $l=0$, the relocated values contribute to the empty samples the $P$-point IDFT of the filled cells,
\begin{equation}
\tilde x_\mathrm{rel}[gL]=\frac{1}{\sqrt{P}}\sum_{q=0}^{N_\mathrm{p}-1}X[k_q]\,e^{j2\pi(q+1)g/P},
\label{eq:fill}
\end{equation}
with an average power over $g$ of $N_\mathrm{p}(L-1)/M\approx N_\mathrm{p}/P$ relative to the data, i.e.\ nearly the data power for $P$ close to $N_\mathrm{p}+1$. The exchange of $X[k_q]$ for $p[q]$ on comb 0 adds by~\eqref{eq:col} the value $(\tilde Z[0,l]-Z[0,l])/\sqrt{P}$ to all samples with time offset $l$, which the receiver removes when restoring comb 0.

\subsection{Subcarrier Mapping, IFFT and Cyclic Prefix}
The values $\tX[k]$ are mapped to the subcarriers $K+k$ of an $N$-point IFFT, and $N_\mathrm{cp}$ cyclic-prefix samples are added:
\begin{equation}
s[n]=\frac{1}{\sqrt{N}}\sum_{k=0}^{M-1}\tX[k]\,e^{j2\pi(K+k)n/N},\ \ n=-N_\mathrm{cp},\dots,N\!-\!1.
\label{eq:ifft}
\end{equation}

\subsection{Optional Peak Cancellation}
The cells $(r,0)$ with $r=N_\mathrm{p}+1,\dots,P-1$ are not used by the relocation. By~\eqref{eq:izak} and~\eqref{eq:col}, a value $a[r]$ in such a cell adds $a[r]/\sqrt{L}$ to comb $r$ and changes only the empty samples, which the receiver discards. These cells can carry a peak-cancelling signal without side information. We use clip-and-project: the excess $c[n]=s[n]-A\,s[n]/|s[n]|$ of the samples $n\ge0$ with $|s[n]|>A$ ($c[n]=0$ otherwise), with spectrum $C[k]=\frac{1}{\sqrt{N}}\sum_{n=0}^{N-1}c[n]\,e^{-j2\pi(K+k)n/N}$ on the allocated subcarriers, is projected onto each cell,
\begin{equation}
a[r]\leftarrow a[r]-\frac{\lambda}{\sqrt{L}}\sum_{t=0}^{L-1}C[r+tP],
\label{eq:tr}
\end{equation}
and~\eqref{eq:ifft} is recomputed, for $N_\mathrm{it}$ iterations, with step size $\lambda$, from $a[r]=0$. The comb sum in~\eqref{eq:tr} is the projection of the excess onto cell $(r,0)$, as in~\eqref{eq:empty}. A larger $P$ gives more unused cells, so the option trades data samples for PAPR, without changing the pilots or the relocated data. It is intended for power-limited UEs, since each iteration costs an FFT pair.
\section{Channel}
\label{sec:ch}

\subsection{Time Domain}
We use the 3GPP TDL model~\cite{3gpp38901,3gpp38104}: path $i$ has a delay of $\tau_i$ samples at the sampling rate $N\Delta f$, with $\Delta f$ the subcarrier spacing, and a complex gain $h_{m,i}$ in symbol $m$ of the slot. The received samples of symbol $m$ are
\begin{equation}
y_m[n]=\sum_i h_{m,i}\,s_m\big[n-\tau_i\big]+v_m[n],
\label{eq:ytime}
\end{equation}
with noise $v_m[n]\sim\mathcal{CN}(0,\sigma^2)$. For the derivation, the delays are on the sampling grid and the gains are constant within a symbol (the simulations use neither assumption). The gains of different paths are independent complex Gaussian (Rayleigh fading), with average powers $\Omega_i=\mathbb{E}\big[|h_{m,i}|^2\big]$ summing to one, and follow the Jakes Doppler spectrum~\cite{3gpp38901,3gpp38104},
\begin{equation}
\mathbb{E}\big[h_{m,i}\,h_{\mu,i}^*\big]=\Omega_i\,\rho[m-\mu],
\label{eq:wssus}
\end{equation}
with the time correlation $\rho[m-\mu]=J_0\big(2\pi f_\mathrm{D}T(m-\mu)\big)$, where $f_\mathrm{D}$ is the maximum Doppler frequency, $T=(N+N_\mathrm{cp})/(N\Delta f)$ is the symbol duration, and $J_0(\cdot)$ is the zeroth-order Bessel function of the first kind.

\subsection{Frequency Domain}
After cyclic-prefix removal, $n\ge0$, and for $\tau_i\le N_\mathrm{cp}$, $s_m[n-\tau_i]=s_m[(n-\tau_i)\bmod N]$ by~\eqref{eq:ifft}. The $N$-point FFT
\begin{equation}
Y_m[k]=\frac{1}{\sqrt{N}}\sum_{n=0}^{N-1}y_m[n]\,e^{-j2\pi(K+k)n/N}
\label{eq:fft}
\end{equation}
turns each cyclic shift into a phase:
\begin{align}
Y_m[k]&=H_m[k]\,\tX_m[k]+V_m[k],\notag\\
H_m[k]&=\sum_ih_{m,i}\,e^{-j2\pi(K+k)\tau_i/N},
\label{eq:rxf}
\end{align}
with $V_m[k]\sim\mathcal{CN}(0,\sigma^2)$. By~\eqref{eq:wssus}, the correlation separates into time and frequency parts,
\begin{equation}
\mathbb{E}\big[H_m[k]H_\mu^*[\kappa]\big]=\rho[m-\mu]\,R[k-\kappa],
\label{eq:Rtf}
\end{equation}
with the frequency correlation at the subcarrier lag $k-\kappa$
\begin{equation}
R[k-\kappa]=\sum_i\Omega_i\,e^{-j2\pi (k-\kappa)\tau_i/N}.
\label{eq:Rf}
\end{equation}
Model~\eqref{eq:rxf} neglects the ICI of the variation within a symbol, whose relative power is about $\sigma_\mathrm{I}^2=(\pi f_\mathrm{D}/\Delta f)^2/3$ for the Jakes spectrum~\cite{wang2006}, and the inter-symbol interference of paths beyond the CP; the simulations include both.

\section{Receiver}
\label{sec:rx}

The receiver undoes the transmitter blocks in reverse order (Fig.~\ref{fig:block}); a hat marks the receiver's estimates.

\subsection{Pilot Observations}
The pilot subcarriers $k_q$ of~\eqref{eq:bins} hold only the pilots, by~\eqref{eq:reloc}. With~\eqref{eq:rxf}, each gives the least-squares (LS) observation
\begin{equation}
\frac{Y_m[k_q]}{p[q]}=H_m[k_q]+\frac{V_m[k_q]}{p[q]},
\label{eq:ls}
\end{equation}
which contains no data. Its noise term has variance $\sigma^2$ since $|p[q]|=1$, and the ICI neglected in~\eqref{eq:rxf} adds approximately $\sigma_\mathrm{I}^2$. Unlike that of a DMRS symbol, this ICI comes from the unknown data of the neighboring subcarriers. Let $\mathcal{S}$ be the set of pilot positions $(\mu,\kappa)$ of the slot, $N_\mathrm{p}$ per symbol.

\subsection{2-D LMMSE Estimation over the Slot}
The channel at every position of the slot is estimated as a weighted sum of all LS observations,
\begin{equation}
\hat H_m[k]=\sum_{(\mu,\kappa)\in\mathcal{S}}W_{m,k}[\mu,\kappa]\,\frac{Y_\mu[\kappa]}{p_\mu[\kappa]},
\label{eq:lmmse}
\end{equation}
where $p_\mu[\kappa]=p[q]$ for $\kappa=k_q$ in symbol $\mu$. By the orthogonality principle~\cite{hoeher1997}, the error $H_m[k]-\hat H_m[k]$ is uncorrelated with every observation, which with~\eqref{eq:ls} and~\eqref{eq:Rtf} gives, for every $(\eta,\vartheta)\in\mathcal{S}$,
\begin{multline}
\sum_{(\mu,\kappa)\in\mathcal{S}}W_{m,k}[\mu,\kappa]\Big(\rho[\mu-\eta]\,R[\kappa-\vartheta]\\
+\Big(\frac{\sigma^2}{|p_\mu[\kappa]|^2}+\sigma_\mathrm{I}^2\Big)\delta[\mu-\eta]\,\delta[\kappa-\vartheta]\Big)=\rho[m-\eta]\,R[k-\vartheta],
\label{eq:wf}
\end{multline}
a set of $|\mathcal{S}|$ linear equations whose matrix is the same for all positions $(m,k)$ and is factorized once per slot configuration. The resulting normalized mean square error (NMSE) is
\begin{equation}
\varepsilon=1-\frac{1}{N_\mathrm{s}M}\sum_{m,k}\ \sum_{(\mu,\kappa)\in\mathcal{S}}W_{m,k}[\mu,\kappa]\,\rho[m-\mu]\,R[\kappa-k],
\label{eq:nmse}
\end{equation}

The pilot spacing is limited by the delay spread. By~\eqref{eq:rxf}, every path is a complex exponential over the subcarriers with $\tau_i/N$ cycles per subcarrier, so on pilots spaced $\Delta k$ subcarriers apart it makes $\Delta k\,\tau_i/N$ cycles per pilot, and paths whose $\Delta k\,\tau_i/N$ differ by an integer alias. In one symbol $\Delta k=M/N_\mathrm{p}$; across the $N_\mathrm{h}$ symbols of a hopping period the positions interleave to $\Delta k=M/(N_\mathrm{p}N_\mathrm{h})$. All delays in $[0,\max_i\tau_i]$ are separated, i.e.\ the channel is sampled above its Nyquist rate in frequency, if
\begin{equation}
\frac{M}{N_\mathrm{p}N_\mathrm{h}}\,\max_i \tau_i<N,
\label{eq:Pmax}
\end{equation}
that is, the pilot spacing of a hopping period in hertz, $M\Delta f/(N_\mathrm{p}N_\mathrm{h})$, times the largest delay in seconds, $\max_i\tau_i/(N\Delta f)$, is below one. Equivalently, $N_\mathrm{p}N_\mathrm{h}$ must exceed the time--bandwidth product $M\max_i\tau_i/N$ of the allocation, which depends on the bandwidth $M\Delta f$. Hopping relaxes it by the factor $N_\mathrm{h}$ at no cost in pilots, provided the channel is correlated over $N_\mathrm{h}$ symbols, which~\eqref{eq:wf} weighs through $\rho[\cdot]$. The pilots of a slot form a sheared time--frequency lattice, as scattered pilots do~\cite{dvbt}: its aliases are shifted in delay and, except every $N_\mathrm{h}$-th one, also in Doppler by at least $1/N_\mathrm{h}$ cycles per symbol, so they do not overlap the channel if, besides~\eqref{eq:Pmax}, $2f_\mathrm{D}TN_\mathrm{h}<1$, i.e.\ the hopping period is shorter than half a Doppler period. Beyond these conditions, the accuracy grows with the pilots per slot.

\subsection{MMSE Equalization}
Each subcarrier is MMSE-equalized~\cite{falconer2002},
\begin{equation}
\hat X_m[k]=\frac{1}{\beta_m}\,\frac{\hat H_m^*[k]}{|\hat H_m[k]|^2+\sigma^2+\sigma_\mathrm{I}^2}\,Y_m[k],
\label{eq:eq}
\end{equation}
where $\beta_m$ is the mean of $|\hat H_m[k]|^2/(|\hat H_m[k]|^2+\sigma^2+\sigma_\mathrm{I}^2)$ over $k$, which removes the bias of the MMSE coefficient after the IDFT (exactly for $\hat H_m=H_m$), and $\hat X_m[k]$ estimates $\tX_m[k]$. 

\subsection{Reading the Relocated Values}
By~\eqref{eq:reloc} and~\eqref{eq:empty}, comb $q+1$ of $\tX$ sums to $\sqrt{L}\,X[k_q]$. The receiver reads the estimates $\hat z_q$ of $X[k_q]$ as comb sums,
\begin{equation}
\hat z_q=\frac{1}{\sqrt{L}}\sum_{t=0}^{L-1}\hat X[(q+1)+tP],
\label{eq:read}
\end{equation}
and writes them back on the pilot subcarriers,
\begin{equation}
\hat X[k_q]\leftarrow\hat z_q.
\label{eq:restore}
\end{equation}
For a noise-free, perfectly equalized symbol, $\hat X=\tX$, so~\eqref{eq:read} gives $\hat z_q=X[k_q]$, and after~\eqref{eq:restore} $\hat X$ differs from $X$ only by the constants $X[k_q]/\sqrt{L}$ on combs $1,\dots,N_\mathrm{p}$ and by the peak-cancelling values on the unused combs. With the MMSE coefficients of~\eqref{eq:eq} or an imperfect estimate, the end-to-end gain varies over the comb, so $\hat z_q$ contains a small residual of the data of comb $q+1$ and the constants leak slightly onto the data samples, like the residual inter-symbol interference of plain DFT-s-OFDM.

\subsection{IDFT and Detection}
The IDFT~\eqref{eq:idft} applied to $\hat X[k]$ gives the estimates $\hat x[l+gL]$. Since $e^{j2\pi(r+tP)(l+gL)/M}=e^{j2\pi r(l+gL)/M}e^{j2\pi tl/L}$, a constant $a/\sqrt{L}$ on comb $r$ contributes to $\hat x[l+gL]$
\begin{equation}
\frac{a\,e^{j2\pi r(l+gL)/M}}{\sqrt{ML}}\sum_{t=0}^{L-1}e^{j2\pi tl/L},
\label{eq:leftover}
\end{equation}
which is non-zero only for $l=0$, i.e.\ on the samples $\hat x[gL]$. The peak-canceling values thus fall on the empty samples, which the receiver discards, and the data positions give $\hat x[l+gL]=x[l+gL]$, $l\neq0$, for a noise-free symbol, without side information. The data estimates $\hat d[\cdot]$ are the samples $\hat x[l+gL]$, $l\neq0$, in increasing order of $l+gL$.

The read-back~\eqref{eq:read} is linear and precedes detection, so no decision error can propagate. It combines the equalized noise of the $L$ subcarriers of comb $q+1$ with weight $1/\sqrt{L}$ each, so $\hat z_q$ has the average equalized noise variance of comb $q+1$, like the data value it replaces, and since comb $q+1$ spans the allocation, it has the same frequency diversity. The receiver is therefore the standard DFT-s-OFDM receiver with $N_\mathrm{p}$ comb sums inserted before the IDFT.

\section{Overhead, Complexity and NR Integration}
\label{sec:nr}

\subsection{Overhead}
A symbol with $N_\mathrm{p}$ pilots carries $M-P$ data symbols, against $M$ for plain DFT-s-OFDM and none for a DMRS symbol. With pilots in all $N_\mathrm{s}$ symbols of a slot, the slot carries $N_\mathrm{s}(M-P)$ data symbols, against $(N_\mathrm{s}-N_\mathrm{DMRS})M$ for NR with $N_\mathrm{DMRS}$  symbols, so it carries more data whenever $N_\mathrm{s}P<N_\mathrm{DMRS}M$. For a fixed delay spread,~\eqref{eq:Pmax} bounds the pilot spacing in subcarriers, so for a fixed $N_\mathrm{h}$, $N_\mathrm{p}$ grows in proportion to $M$ and the overhead $P/M$ tends to a constant, like the $1/N_\mathrm{s}$ of a DMRS symbol.

\subsection{Complexity}
The transmitter adds $N_\mathrm{p}$ values to $L$ subcarriers each in~\eqref{eq:reloc}, i.e.\ $N_\mathrm{p}L=N_\mathrm{p}M/P<M$ complex additions, and the scaling by $1/\sqrt{L}$ can be merged into the read-out of $X[k_q]$. The receiver needs $N_\mathrm{p}L$ additions for~\eqref{eq:read} and $N_\mathrm{p}$ assignments for~\eqref{eq:restore}. Both are negligible against the $M$-point DFT. Each iteration of the optional peak cancellation~\eqref{eq:tr} needs one $N$-point FFT and IFFT and $2(P-1-N_\mathrm{p})L$ additions. The estimator~\eqref{eq:wf} has the size of the number of pilots per slot, $N_\mathrm{s}N_\mathrm{p}$, against $N_\mathrm{DMRS}M/2$ for $N_\mathrm{DMRS}$ DMRS symbols of configuration type~1. 

\subsection{Integration into NR}
The DFT size, the resource allocation, the subcarrier mapping, the IFFT and the cyclic prefix of transform-precoded PUSCH are unchanged. The data mapping, however, leaves $P$ samples empty, which changes the rate matching, and the relocation is a new block after the DFT. The scheme is described by the number of pilots $N_\mathrm{p}$ and the hopping period $N_\mathrm{h}$, which can be configured semi-statically like the DMRS additional positions~\cite{3gpp38211}; $P$ follows from $M$, $N_\mathrm{p}$ and $N_\mathrm{h}$ by the rule of Section~\ref{sec:tx}-D. The DMRS symbols of a slot can then be removed or reduced to the front-loaded one, and the PT-RS can be dropped since the pilots can track the phase in every symbol. The base station (gNB) adds the pilot extraction to its channel estimator and the comb sums~\eqref{eq:read} before the IDFT. FDSS multiplies $\tX[k]$ by a known window, which the receiver removes before~\eqref{eq:read}; $\pi/2$-BPSK applies to the data symbols $d[\cdot]$, but the filled empty samples interrupt its alternating phase, so part of its PAPR advantage is lost. Co-scheduled UEs or layers need disjoint pilot subcarriers, e.g., different pilot positions on comb 0, at the cost of a larger $P$.

\section{Simulation Results}
\label{sec:sim}

\begin{table}[t]
\centering
\caption{Simulation Parameters}
\label{tab:params}
\footnotesize
\setlength{\tabcolsep}{3pt}
\begin{tabular}{@{}p{2.1cm}p{6.2cm}@{}}
\toprule
$M$, $N$, $\Delta f$ & 192 (16 resource blocks); 512 (link), $8M$ (PAPR); 30~kHz; carrier 3.5~GHz \\
$N_\mathrm{cp}$, $N_\mathrm{s}$ & 36 (2.34~\textmu s), 14 (0.5~ms slot) \\
Proposed & $(N_\mathrm{p},P)=(2,3)$, $(4,6)$, $(8,12)$; ZC pilots, $u=1$; hopping $N_\mathrm{h}=4$ for $N_\mathrm{p}=2,4$, $N_\mathrm{h}=2$ for $N_\mathrm{p}=8$, and $N_\mathrm{h}=1$ for reference \\
Benchmarks & NR DMRS type~1 with 1 to 4 symbols~\cite{3gpp38211}; pre-DFT RS~\cite{gudimitla2024}: ZC reference signal of 13 samples, cyclic prefix and suffix of 6 samples; puncturing: $(N_\mathrm{p},N_\mathrm{h})=(4,4)$, $(8,2)$, pilots as in the proposed scheme \\
Peak cancellation & $A=1.6$ times the RMS of $s[n]$, $\lambda=4$, $N_\mathrm{it}=30$ \\
Channels & TDLB100, TDLC300~\cite{3gpp38104}; 3, 60, 120, 250, 350, 500~km/h ($f_\mathrm{D}\le1.62$~kHz); Rayleigh fading, Jakes spectrum, sum of 16 sinusoids per path~\cite{zheng2003} \\
Monte Carlo & $10^5$ symbols (PAPR), 600 independent slots per point (link) \\
\bottomrule
\end{tabular}
\end{table}

\begin{table}[!t]
\centering
\caption{Slot Overhead and Complexity}
\label{tab:overhead}
\footnotesize
\setlength{\tabcolsep}{3pt}
\begin{tabular}{@{}lcccc@{}}
\toprule
Scheme & Data sym- & Data & Pilots per slot & Additions \\
 & bols per slot & fraction & (estim. size) & per symbol \\
\midrule
NR, 1 DMRS symbol & 2496 & 92.9\% & 96 & -- \\
NR, 2 DMRS symbols & 2304 & 85.7\% & 192 & -- \\
NR, 3 DMRS symbols & 2112 & 78.6\% & 288 & -- \\
NR, 4 DMRS symbols & 1920 & 71.4\% & 384 & -- \\
Pre-DFT RS~\cite{gudimitla2024} & 2338 & 87.0\% & 182 & -- \\
Punctured, $N_\mathrm{p}=4$ & 2688 & 100\% & 56 & -- \\
Punctured, $N_\mathrm{p}=8$ & 2688 & 100\% & 112 & -- \\
Proposed, $N_\mathrm{p}=2$ & 2646 & 98.4\% & 28 & 128 \\
Proposed, $N_\mathrm{p}=4$ & 2604 & 96.9\% & 56 & 128 \\
Proposed, $N_\mathrm{p}=8$ & 2520 & 93.8\% & 112 & 128 \\
\bottomrule
\end{tabular}

\vspace{2pt}
\end{table}

In this section, the performance of the proposed in-symbol pilots is evaluated in terms of the PAPR, the normalized mean square error (NMSE) of the channel estimate, the BER and the effective spectral efficiency, and is compared with the NR DMRS and with two in-symbol pilot designs reviewed in Section~I. Unless otherwise stated, the simulation parameters are those listed in Table~\ref{tab:params}. The allocation spans $M=192$ subcarriers (16 resource blocks), an NR DFT size, and the proposed scheme uses the smallest grid of Section~\ref{sec:tx}-D, i.e.\ $N_\mathrm{p}=2$, 4 and 8 pilots with $P=3$, 6 and 12, which leaves 189, 186 and 180 data symbols per DFT-s-OFDM symbol. The resulting slot overhead is summarized in Table~\ref{tab:overhead}.

The link-level results are obtained over the 3GPP delay profiles TDLB100 and TDLC300~\cite[Annex~G]{3gpp38104}, whose root-mean-square (RMS) delay spreads are 100 and 300~ns and whose largest delays are 0.48 and 2.60~\textmu s. Each path is applied as a band-limited fractional delay with a gain that varies within the symbol, so that ICI and inter-symbol interference (the last TDLC300 path exceeds the cyclic prefix) affect every scheme. An independent channel realization is drawn for each slot, and all schemes are evaluated over the same realizations, with the equalizer~\eqref{eq:eq}, one receive antenna and uncoded 16 quadrature amplitude modulation (16-QAM); the peak cancellation is not used in the link-level results.

Three benchmarks are considered. The first is the NR DMRS of configuration type~1 (every second subcarrier, no data in the DMRS symbols) with 1, 2, 3 or 4 single-symbol DMRS at the positions $\{2\}$, $\{2,11\}$, $\{2,7,11\}$ and $\{2,5,8,11\}$ of PUSCH mapping type~A~\cite{3gpp38211}, whereas the proposed slot carries pilots in all 14 symbols and no DMRS symbol. The second is the pre-DFT RS of~\cite{gudimitla2024}, in which a ZC reference signal of $1+M/16=13$ samples, extended by a cyclic prefix and suffix of 6 samples each, precedes the DFT in every symbol and leaves 167 data symbols; the 13-tap impulse response of each symbol is estimated by least squares from the received block and transformed to the $M$ subcarriers by a DFT. The third is puncturing, where $N_\mathrm{p}=4$ or 8 subcarriers of the DFT output are overwritten with the pilots of the proposed scheme, i.e.\ with the same positions, hopping, ZC values and power. The punctured symbols carry all $M$ data symbols but lose their components on the pilot subcarriers; by~\eqref{eq:idft}, this adds to every data symbol an error of power $N_\mathrm{p}/M$, which bounds the ratio of its power to the error power by $M/N_\mathrm{p}$, i.e.\ 16.8 and 13.8~dB. Hence, the proposed and punctured slots differ only in how the data on the pilot subcarriers are treated. For a fair comparison, the proposed and punctured slots share the estimator~\eqref{eq:lmmse}--\eqref{eq:wf} with the true delay profile, $f_\mathrm{D}$, $\sigma^2$ and $\sigma_\mathrm{I}^2$.

The PAPR of a symbol $s[n]$ of~\eqref{eq:ifft} is $\max_n|s[n]|^2/\big(\frac{1}{N}\sum_n|s[n]|^2\big)$ over the samples. It is evaluated with $N=8M$, so that the samples approximate the continuous-time PAPR~\cite{wunder2013}, and is characterized by its complementary cumulative distribution function (CCDF), $\Pr(\mathrm{PAPR}>\gamma)$, and by $\gamma_{10^{-3}}$, the level exceeded with probability $10^{-3}$; OFDM with QAM symbols on the same $M$ subcarriers serves as a reference. The SNR is defined as $E_\mathrm{s}/N_0=1/\sigma^2$ for unit average energy per data or DMRS resource element, and the proposed symbols carry the energy $(M-P+N_\mathrm{p})/M$ per element by~\eqref{eq:energy}. To account for both the overhead and the estimation accuracy, the effective spectral efficiency of a slot is defined as the data fraction of Table~\ref{tab:overhead} times $I(1/\xi)$, where $\xi$ is the mean square error of the detected data symbols of the slot and $I(\cdot)$ is the mutual information of 16-QAM on a Gaussian channel at the given SNR.

\begin{figure}[!t]
\centering
\includegraphics[width=\columnwidth]{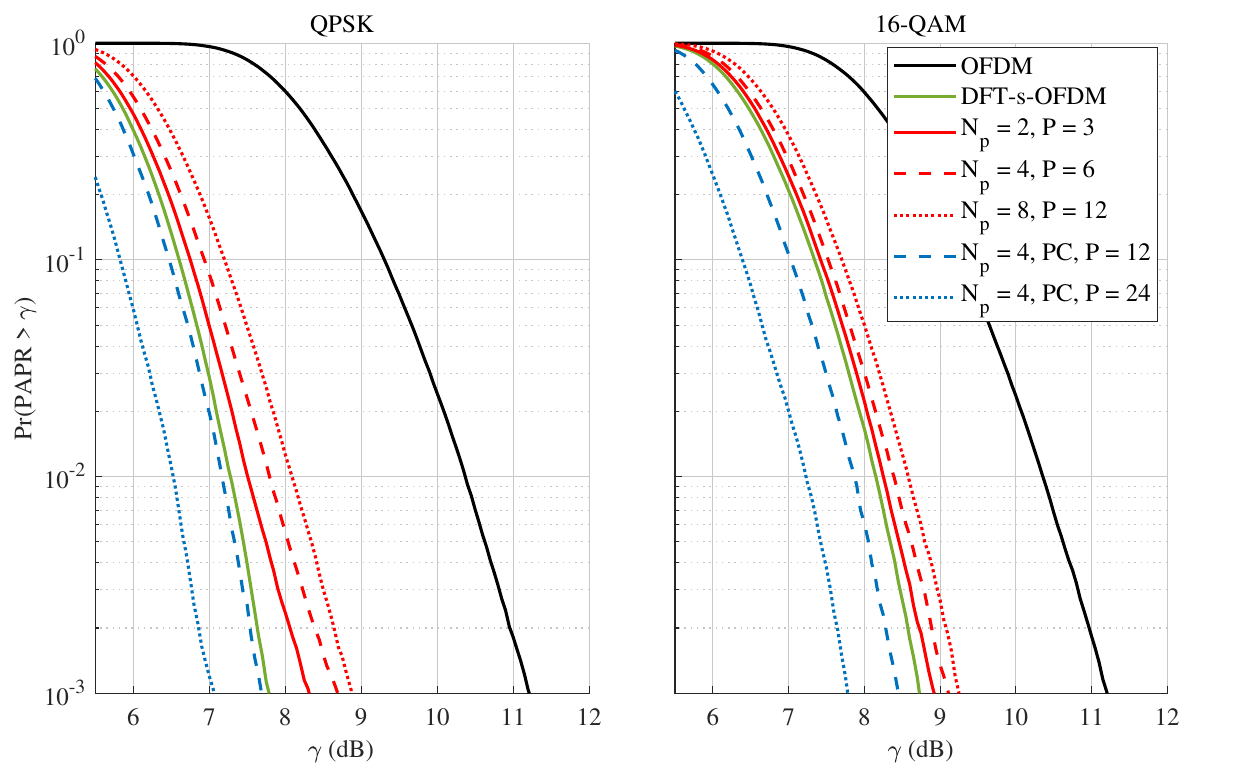}
\caption{PAPR CCDF for QPSK (left) and 16-QAM (right).}
\label{fig:ccdf}
\end{figure}

\subsection{PAPR Performance}
Fig.~\ref{fig:ccdf} depicts the CCDF of the PAPR for quadrature phase-shift keying (QPSK) and 16-QAM. As expected, $\gamma_{10^{-3}}$ of plain DFT-s-OFDM is 3.4 and 2.5~dB below that of OFDM for QPSK and 16-QAM, respectively. The proposed symbols remain close to plain DFT-s-OFDM: with $N_\mathrm{p}=2$, 4 and 8 pilots, $\gamma_{10^{-3}}$ increases by 0.5, 0.9 and 1.0~dB for QPSK and by 0.2, 0.4 and 0.5~dB for 16-QAM, so that most of the PAPR advantage of DFT-s-OFDM over OFDM is retained. The increase grows with $N_\mathrm{p}$, since the pilots and the relocated values form a larger part of the symbol. For QPSK, an increase of about one decibel with four or eight pilots is not negligible for a UE at maximum power, which motivates the optional peak cancellation (PC). The ZC pilots also contribute to the low PAPR, owing to the even spreading predicted by~\eqref{eq:pilotrow}, and their effect grows with $N_\mathrm{p}$. With PC on the unused cells, $N_\mathrm{p}=4$ pilots with $P=12$, i.e.\ six data symbols fewer than with $P=6$, are about 0.1~dB below plain DFT-s-OFDM for QPSK (7.7 against 7.8~dB) and 0.3~dB below it for 16-QAM; with $P=24$, they are 0.7 and 1.0~dB below it, at the cost of at most 0.13 and 0.24~dB of additional power.

\begin{figure*}[t]
\centering
\subfloat[TDLB$100$]{\includegraphics[width=0.245\textwidth]{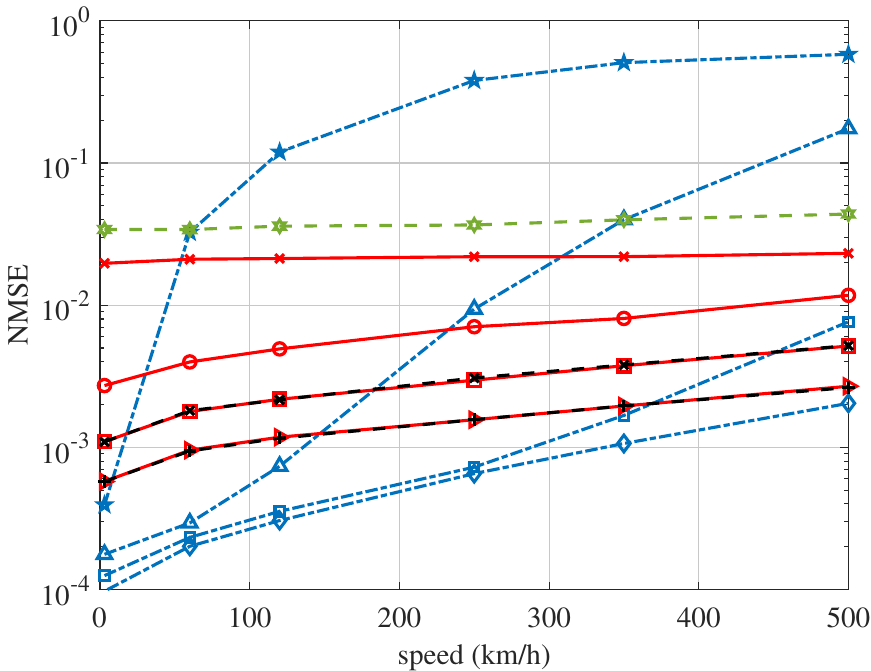}\label{fig:NMSE_TDLB100}}
\subfloat[TDLC$300$]{\includegraphics[width=0.245\textwidth]{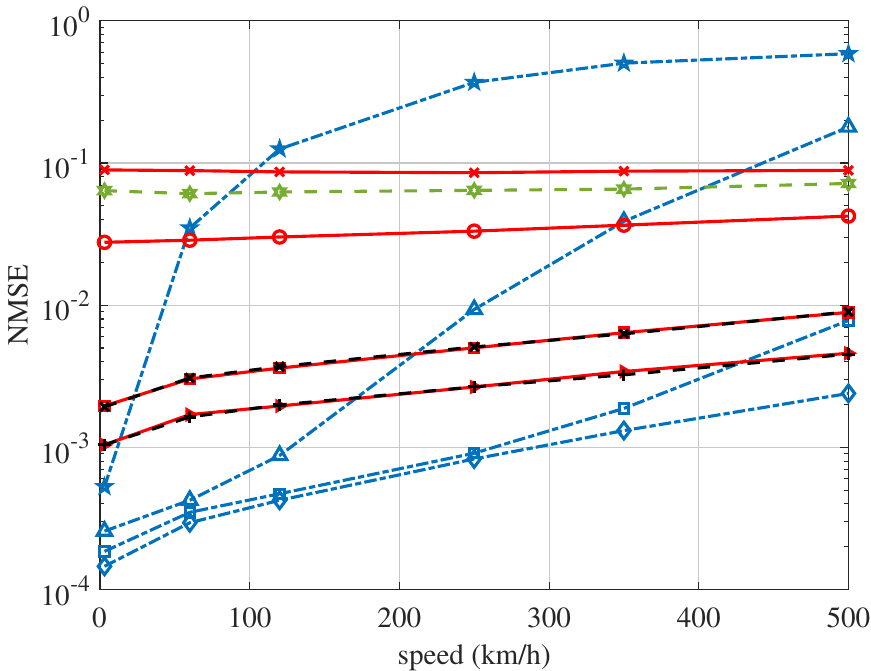}\label{fig:NMSE_TDLC300}}
\subfloat[TDLB$100$]{\includegraphics[width=0.245\textwidth]{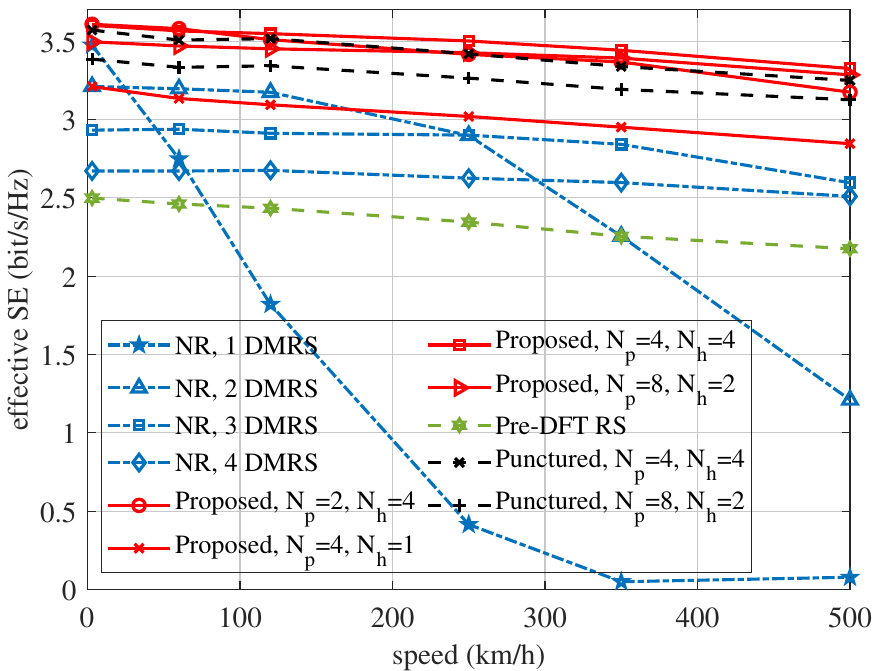}\label{fig:SE_TDLB100}}
\subfloat[TDLC$300$]{\includegraphics[width=0.245\textwidth]{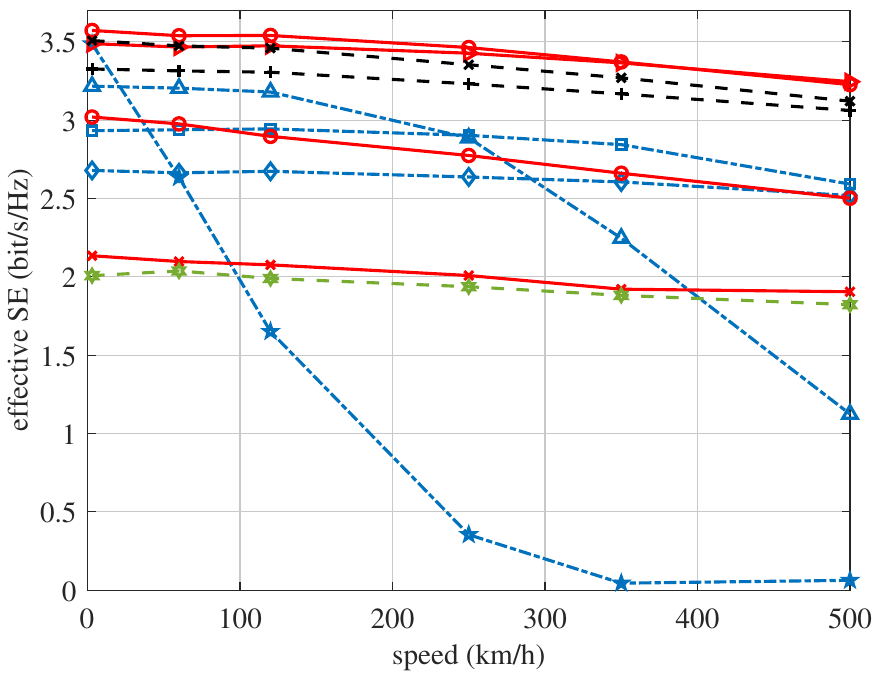}\label{fig:SE_TDLC300}}
\caption{NMSE and effective spectral efficiency versus speed at $E_\mathrm{s}/N_0=20$~dB.}
\label{fig:speed}
\end{figure*}

\subsection{Channel Estimation Accuracy}
Figs.~\ref{fig:speed}(a) and~\ref{fig:speed}(b) show the NMSE versus the UE speed at $E_\mathrm{s}/N_0=20$~dB. With a single DMRS symbol, the NMSE exceeds $3\times10^{-2}$ from 60~km/h, since the channel must be extrapolated over up to eleven symbols, whereas additional DMRS symbols keep it low at the price of the data fraction. Since the proposed slot carries pilots in every symbol, its NMSE grows slowly with the speed: on TDLB100, it increases from $1.1\times10^{-3}$ to $5.2\times10^{-3}$ for $N_\mathrm{p}=4$ with hopping and from $5.8\times10^{-4}$ to $2.7\times10^{-3}$ for $N_\mathrm{p}=8$, the latter being close to four DMRS symbols at 500~km/h. This NMSE is limited by the small number of pilots per slot and by the ICI on the pilots.

The hopping is essential. Without it, $N_\mathrm{p}=4$ pilots spaced $M/N_\mathrm{p}=48$ subcarriers apart saturate at an NMSE of $9\times10^{-2}$ on TDLC300, whose largest delay violates~\eqref{eq:Pmax}, and at $2\times10^{-2}$ even on TDLB100, which satisfies it, because four pilots per symbol leave an interpolation error and the band edges to extrapolation. With $N_\mathrm{h}=4$, the positions interleave to a spacing of 12 subcarriers, and~\eqref{eq:Pmax} holds for both profiles. Likewise, $N_\mathrm{p}=2$ pilots with $N_\mathrm{h}=4$ (spacing 24) suffice for TDLB100 but not for TDLC300.

Regarding the benchmarks, the punctured pilots attain nearly the same NMSE as the proposed pilots, as both leave the pilot subcarriers free of data. In contrast, the NMSE of the pre-DFT RS lies between $3.4\times10^{-2}$ and $4.4\times10^{-2}$ on TDLB100 and between $6.1\times10^{-2}$ and $7.2\times10^{-2}$ on TDLC300, and is nearly independent of the speed, because each symbol is estimated from its own reference signal. It remains above that of the proposed pilots, since each estimate relies on the 13 samples of one symbol, without the averaging over the slot of the two-dimensional estimator; moreover, on TDLC300, the largest delay, about 15 pre-DFT samples, exceeds the length of the reference signal.

\subsection{Spectral Efficiency and BER}
Figs.~\ref{fig:speed}(c) and~\ref{fig:speed}(d) present the effective spectral efficiency versus the UE speed at $E_\mathrm{s}/N_0=20$~dB. For NR, the trade-off between accuracy and overhead is evident: at 3~km/h, four DMRS symbols yield 23\% less effective spectral efficiency than one, whereas one DMRS symbol collapses at high speed, so that the best NR configuration changes from one DMRS symbol at low speed to three at 500~km/h. In contrast, a proposed configuration with hopping provides the highest effective spectral efficiency at every speed on both profiles. On TDLB100 at low speed, where two pilots per symbol suffice, $N_\mathrm{p}=2$ performs best owing to its 98.4\% data fraction; otherwise $N_\mathrm{p}=4$ performs best, except at 500~km/h on TDLC300, where $N_\mathrm{p}=8$ does. Since $N_\mathrm{p}=4$ and $N_\mathrm{p}=8$ never differ by more than 0.11~bit/s/Hz, a single configuration can serve all speeds without reconfiguration. At 3~km/h, $N_\mathrm{p}=4$ exceeds the best NR configuration, one DMRS symbol, by 0.09 to 0.13~bit/s/Hz, and at 500~km/h, the proposed slots reach 3.33 and 3.25~bit/s/Hz on TDLB100 and TDLC300, against 2.60 and 2.59~bit/s/Hz for the best NR configuration, i.e.\ 28\% and 25\% more.

Among the benchmarks, the punctured slots come closest to the proposed ones. Although they carry all $M$ data symbols, they remain 0.03 to 0.11~bit/s/Hz below the proposed slots with $N_\mathrm{p}=4$ and 0.11 to 0.20~bit/s/Hz below with $N_\mathrm{p}=8$, as the error of the lost components grows with $N_\mathrm{p}$. The pre-DFT RS achieves between 2.18 and 2.50~bit/s/Hz on TDLB100 and between 1.82 and 2.04~bit/s/Hz on TDLC300, below four DMRS symbols at every speed, as its higher NMSE outweighs its data fraction of 87.0\%.

\begin{figure}[!t]
\centering
\includegraphics[width=\columnwidth]{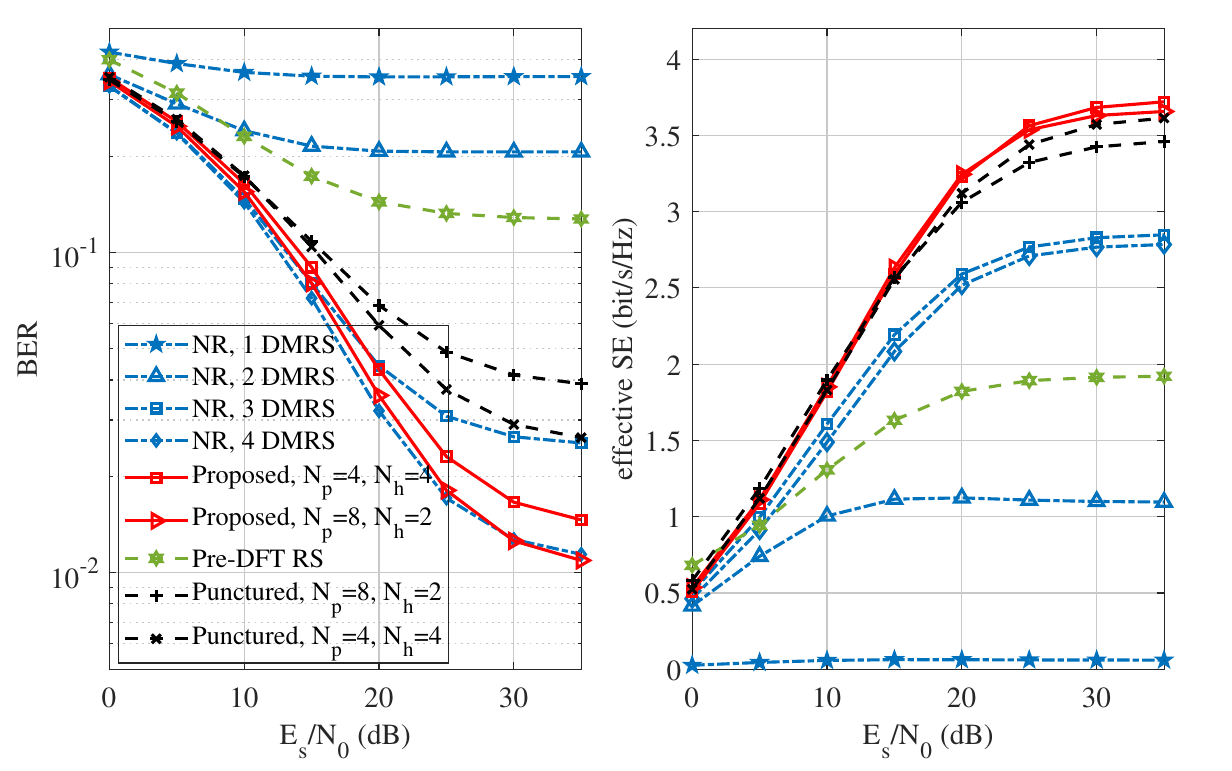}
\caption{BER and effective spectral efficiency versus SNR at 500~km/h on TDLC300.}
\label{fig:snr}
\end{figure}

Fig.~\ref{fig:snr} shows the BER and the effective spectral efficiency versus the SNR at 500~km/h on TDLC300, for the configurations that satisfy~\eqref{eq:Pmax} and for the benchmarks. With hopping, the BER of $N_\mathrm{p}=8$ nearly coincides with that of four DMRS symbols, the most robust NR configuration, while carrying 31\% more data symbols per slot, and $N_\mathrm{p}=4$ requires about 2~dB more SNR to reach a BER of $5\times10^{-2}$ while carrying 36\% more. Below 20~dB, the BER of the proposed configurations is slightly higher than that of four DMRS symbols, as fewer pilot observations are available per slot; their gain in this region therefore stems from the data fraction. Above 25~dB, the NR and proposed slots exhibit an error floor, caused by the ICI at this speed and by the frequency-selective fading of a single receive antenna. The benchmarks saturate at higher BER levels: $2.6\times10^{-2}$ and $3.9\times10^{-2}$ for the punctured slots with $N_\mathrm{p}=4$ and 8, against $1.5\times10^{-2}$ and $1.1\times10^{-2}$ for the proposed slots, and $1.3\times10^{-1}$ for the pre-DFT RS.

In terms of effective spectral efficiency, both proposed configurations outperform every NR configuration and the pre-DFT RS from 5~dB on, by up to 0.9~bit/s/Hz over NR at high SNR, while at 0~dB the pre-DFT RS attains the highest value, 0.68~bit/s/Hz. The comparison with puncturing depends on the SNR. Up to 10~dB, where the noise dominates the error of the lost components, the punctured slots are up to 0.07~bit/s/Hz higher owing to their 100\% data fraction, whereas from 15~dB on, they fall below the proposed slots and reach 3.61 and 3.46~bit/s/Hz at 35~dB, against 3.72 and 3.66~bit/s/Hz for $N_\mathrm{p}=4$ and 8.

\section{Conclusion}
\label{sec:Conc}
This paper has proposed interference-free pilots inside DFT-s-OFDM data symbols, enabling frequency-domain channel estimation in every symbol. Using the Zak transform, we have shown that leaving one stride class of pre-DFT samples empty frees one column of the Zak grid, into which the data of the pilot subcarriers are relocated by adding constants over a few combs. This preserves the energy and the NR DFT sizes, costs as few as one data sample beyond the pilots, and requires only additions. Few pilots per symbol, hopped across symbols, keep the PAPR close to that of plain DFT-s-OFDM while a two-dimensional estimator resolves long delay spreads. On 3GPP channels with uncoded 16-QAM, the proposed slots achieved a higher effective spectral efficiency than the NR DMRS, puncturing and pre-DFT reference signals at 20~dB over the evaluated speeds, by 25\% to 28\% over NR at 500~km/h. Coded, multi-antenna and multi-user evaluations, and estimation under mismatched channel statistics, remain to be studied.


\end{document}